\documentclass[a4paper,aps,prd,10pt,preprintnumbers,showpacs,twocolumn,superscriptaddress,nofootinbib,amsmath,amssymb,preprint]{revtex4}
\usepackage{graphicx,cmap,amsmath}
\usepackage[utf8]{inputenc}
\usepackage[T1]{fontenc}

\def\K{{\cal K}}
\def\Order#1{{\cal O}\left(#1\right)}

\begin{document}

\title{Gravitational quasinormal modes of the Hayward spacetime}

\author{S. V. Bolokhov}\email{bolokhov-sv@rudn.ru}
\affiliation{Peoples' Friendship University of Russia (RUDN University), 6 Miklukho-Maklaya Street, Moscow, 117198, Russia}
\author{Milena Skvortsova}\email{milenas577@mail.ru}
\affiliation{Peoples' Friendship University of Russia (RUDN University), 6 Miklukho-Maklaya Street, Moscow, 117198, Russia}

\begin{abstract}
We study gravitational quasinormal modes of the Hayward spacetime, a regular black-hole geometry that also admits an interpretation as an effective quantum-corrected solution within asymptotically safe gravity. Using both the higher-order WKB method supplemented with Padé approximants and time-domain integration with Prony analysis, we obtain accurate spectra for axial perturbations and explore the impact of the quantum parameter $\gamma$. We find that increasing $\gamma$ systematically raises the oscillation frequencies while reducing the damping rates, making the ringdown longer lived. For the first overtone, the effect of $\gamma$ is noticeably stronger than for the fundamental mode, providing an indication of the so-called “outburst of overtones” previously observed for test fields, and pointing to the particular sensitivity of subdominant modes to near-horizon quantum corrections. In addition, the analytic approximation for quasinormal modes are obtained in the form of expansion beyond the eikonal limit 
\end{abstract}

\maketitle

\section{Introduction}

Quasinormal modes (QNMs) play a central role in black-hole physics, as they govern the characteristic oscillations of spacetime in response to perturbations \cite{Kokkotas:1999bd,Berti:2009kk,Konoplya:2011qq,Bolokhov:2025uxz}. These damped oscillations dominate the ringdown phase of gravitational-wave signals and thus provide a direct observational window into the properties of astrophysical black holes. Beyond their relevance for gravitational-wave astronomy, QNMs also encode fundamental information about the stability of black hole spacetimes and the nature of the underlying gravitational theory. Since the spectrum of QNMs is uniquely determined by the background geometry and the type of perturbation, their study serves as a powerful diagnostic tool for testing classical general relativity and probing possible quantum or higher-curvature corrections. In this way, QNMs establish a direct connection between theoretical models of gravity and measurable astrophysical phenomena, making them an indispensable ingredient in the ongoing effort to understand black holes at both the observational and theoretical levels.

The Hayward spacetime occupies a particularly important place among models of regular black holes. Originally proposed as a toy model of an evaporating black hole with a de Sitter core replacing the central singularity \cite{Hayward:2005gi}, it offers a simple yet consistent framework for addressing the problem of singularities within classical general relativity. More recently, it has been shown that, upon a suitable redefinition of constants, the same metric arises naturally in the context of asymptotically safe gravity, where it represents an effective quantum-corrected black hole solution \cite{Held:2019xde}. This dual interpretation makes the Hayward spacetime an ideal testbed: on the one hand, it is a widely studied representative of regular black holes, while on the other, it provides a window into possible quantum-gravity effects. As such, studying its dynamical properties, including quasinormal spectra and grey-body factors, allows one to probe both classical regularization mechanisms and quantum corrections to black-hole physics within a unified framework.

Although the quasinormal spectra of test fields in the Hayward spacetime have been investigated in \cite{Konoplya:2022hll,Malik:2024tuf}, the case of gravitational perturbations has not yet been systematically analyzed. The only related studies have been performed in the framework of Effective Field Theory \cite{Konoplya:2023ppx,Takahashi:2019oxz}, where both the form of the perturbation equations and the theoretical background differ substantially from the Asymptotic Safety scenario. Thus, a dedicated analysis of gravitational perturbations in the Hayward geometry remains an open and important problem.

In order to address this problem, we shall determine the quasinormal modes of gravitational perturbations in the Hayward spacetime by employing two complementary approaches. First, we use the higher-order WKB method supplemented with Padé approximants, which is known to provide accurate results for low-lying modes in a wide range of black-hole geometries. Second, we verify these results through numerical time-domain integration, which offers an independent check and is particularly reliable in regimes where semi-analytical methods may lose accuracy. Together, these techniques allow for a robust determination of the quasinormal spectrum and the assessment of possible quantum-inspired deviations from the Schwarzschild case.

The paper is organized as follows. In Sec.~II we review the Hayward metric and discuss its dual interpretation both as a model of a regular evaporating black hole and as an effective quantum-corrected geometry in the framework of asymptotic safety. Sec.~III presents the formulation of axial gravitational perturbations in this background and the construction of the corresponding wave equation. In Sec.~IV we describe the boundary conditions for quasinormal modes and outline the numerical and semi-analytical methods employed in our analysis, namely the WKB method with Padé resummation and the time-domain evolution with Prony extraction. Sec.~V contains the main results, including the analytic beyond eikonal approximation, numerical data for the spectrum, and a discussion of the role of the quantum parameter $\gamma$. Finally, Sec.~VI summarizes our conclusions and outlines possible directions for future work.

\section{The Hayward Spacetime and Its Interpretations}

A persistent challenge in black-hole physics is the resolution of curvature singularities that inevitably appear in classical solutions of general relativity. Among the earliest and most influential proposals for a singularity-free geometry is the Hayward metric~\cite{Hayward:2005gi}, which provides a simple model of an evaporating black hole endowed with a regular core. The construction belongs to the wider class of so-called \emph{regular black holes}, whose metrics interpolate smoothly between an exterior Schwarzschild-like behavior at large distances and a de Sitter-like core near the center, thereby eliminating the central singularity while preserving a well-defined event horizon.

The line element of the Hayward spacetime is
\begin{equation}
ds^2 = - f(r)\, dt^2 + \frac{dr^2}{f(r)} + r^2 \left( d\theta^2 + \sin^2\theta\, d\phi^2 \right),
\end{equation}
with the metric function
\begin{equation}\label{fr}
f(r) = 1 - \frac{2 M r^2}{r^3 + 2 M l^2}.
\end{equation}
Here $M$ denotes the black-hole mass, and $l$ is a parameter setting the scale at which quantum or non-classical effects are expected to dominate. In the asymptotic region $r \gg (M l^2)^{1/3}$ the metric reduces to the Schwarzschild form $f(r) \approx 1 - 2M/r$, whereas close to the origin it approaches a de Sitter-like behavior $f(r) \approx 1 - r^2/l^2$, thereby providing a regular core of finite curvature. 

Originally, the parameter $l$ was introduced phenomenologically, inspired by the idea that quantum backreaction or exotic matter sources might soften the singularity. However, later developments have shown that the Hayward form can also be understood from a more fundamental perspective. In particular, Held, Gold, and Eichhorn~\cite{Held:2019xde} demonstrated that, up to a redefinition of constants, the same functional form arises as an effective solution within the framework of asymptotically safe gravity. In this approach the metric function can be expressed as
\begin{equation}
f(r) = 1 - \frac{2 r^2 / M^2}{r^3 / M^3 + \gamma},
\end{equation}
where $\gamma$ parametrizes the strength of the quantum corrections. The requirement
\[
\gamma \lesssim \frac{32}{27}
\]
ensures the existence of an event horizon and prevents the emergence of a naked singularity. Here the effective geometry follows from associating the running cutoff scale with curvature invariants such as the Kretschmann scalar. Different prescriptions for fixing the cutoff lead to a family of quantum-corrected black hole metrics, including those of Bonanno and Reuter~\cite{Bonanno:2000ep} and Platania~\cite{Platania:2019kyx}. While quasinormal spectra of test fields have been studied in several of these spacetimes, gravitational perturbations have so far been considered only in a few specific contexts, such as the Bonanno–Reuter effective geometry~\cite{Bolokhov:2025lnt} and black holes obtained from models of quantum collapse~\cite{Bonanno:2023rzk,Shi:2025gst}.

Thus, the Hayward spacetime acquires a dual interpretation: on the one hand, it provides a tractable model of a regular black hole free from central singularities; on the other, it represents an effective quantum-corrected solution within asymptotically safe gravity. This duality makes the Hayward geometry an especially attractive testing ground for probing how quantum-inspired modifications of the gravitational field affect dynamical properties such as quasinormal modes, grey-body factors, and stability.

\section{Gravitational Perturbations}

The analysis of gravitational perturbations in the Hayward background is more delicate than in the Schwarzschild case, since the metric arises as an effective geometry rather than an exact solution of Einstein’s equations with specified matter content. A fully rigorous treatment of perturbations in this setting is therefore nontrivial. Nevertheless, Ashtekar, Olmedo, and Singh \cite{Ashtekar:2018lag,Ashtekar:2018cay} showed that the essential quantum effects can be consistently mimicked within Einstein gravity by introducing an anisotropic fluid stress–energy tensor. This effective description has been successfully applied in a number of contexts, including the study of black-hole perturbations in quantum-corrected spacetimes \cite{Bouhmadi-Lopez:2020oia,Konoplya:2024lch}. Under the simplifying assumption that fluctuations along the anisotropy direction do not contribute in the axial sector, one can treat the dynamics of axial perturbations in close analogy with the standard Regge–Wheeler formalism.  

This approach parallels earlier analyses in which extra fields were present but decoupled from the axial sector, such as black holes coupled to scalar fields \cite{Bronnikov:2012ch,Chen:2019iuo}. In the Regge–Wheeler gauge \cite{Regge:1957td}, the nonvanishing components of the axial perturbations $h_{\mu\nu}$ are parametrized by two functions $h_0(t,r)$ and $h_1(t,r)$, multiplied by the appropriate angular dependence through Legendre polynomials. Substituting this ansatz into Einstein’s equations with the anisotropic fluid source and imposing the condition that anisotropy perturbations vanish, one obtains a coupled system for $h_0$ and $h_1$. Eliminating one of the variables and performing the usual change of coordinates $dr^* = dr / f(r)$, together with a rescaling of the perturbation function, the system reduces to a Schrödinger-like wave equation,  
\begin{equation}\label{wavelike}
\frac{d^2\Psi}{dr_{*}^2} + \left(\omega^2 - V(r)\right)\Psi = 0,
\end{equation}
with the effective axial potential
\begin{equation}
V(r) = f(r)\left(\frac{2 f(r)}{r^2} - \frac{f'(r)}{r} + \frac{(\ell+2)(\ell-1)}{r^2}\right),
\end{equation}
where $\ell=2,3,\ldots$ is the multipole number.  

Similar simplifications—neglecting certain sector-specific perturbations—have been employed in various related studies \cite{Berti:2003yr,Kokkotas:1993ef,Konoplya:2006ar}. While such assumptions may omit possible corrections to the complete gravitational spectrum, they are expected to provide a reliable first approximation as long as deviations from the Schwarzschild geometry remain moderate. This is consistent with the perturbative nature of quantum corrections, which typically affect the background only slightly.

\begin{figure}
\resizebox{\linewidth}{!}{\includegraphics{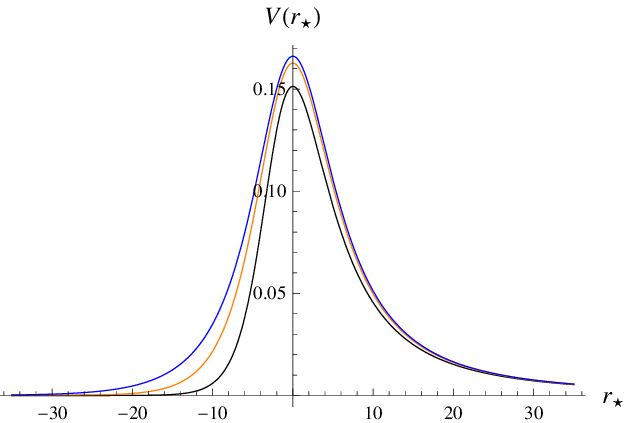}}
\caption{Potential as a function of the tortoise coordinate for $\ell=2$ axial gravitational perturbations: $\gamma=0$ (black)  (green) $\gamma=0.9$ (orange), and $\gamma=1.1$ (blue); $M=1$.}\label{fig:pot1}
\end{figure}

\begin{figure}
\resizebox{\linewidth}{!}{\includegraphics{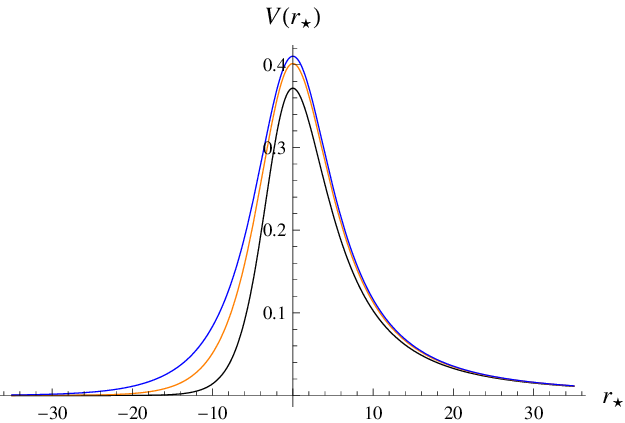}}
\caption{Potential as a function of the tortoise coordinate for $\ell=2$ axial gravitational perturbations: $\gamma=0$ (black)  $\gamma=0.9$ (orange), and $\gamma=1.1$ (blue); $M=1$.}\label{fig:pot2}
\end{figure}

The effective potentials for axial gravitational perturbations of the Hayward spacetime exhibit the familiar single-barrier shape characteristic of black-hole backgrounds. Increasing the quantum correction parameter $\gamma$ raises the height of the potential and slightly shifts its maximum toward the horizon\footnote{The position of the potential maximum moves closer to the horizon when plotted in the radial coordinate $r$, while in the tortoise coordinate $r^{*}$ this effect is not visible, as the additive constant can be chosen such that the peak always lies at $r^{*}=0$.}, while at larger distances the potential rapidly approaches the Schwarzschild form. As a result, quantum effects are confined mainly to the near-horizon region, where they enhance the barrier and thereby suppress transmission probabilities. This behavior is consistent across different multipole numbers, with the overall structure of the potential remaining qualitatively similar to the Schwarzschild case.

\section{Quasinormal Modes: Boundary Conditions and Methods}

Quasinormal modes of massless perturbations are obtained as solutions of the master wave equation \ref{wavelike}. The physically motivated boundary conditions for asymptotically flat spacetimes require purely ingoing waves at the event horizon and purely outgoing waves at infinity:
\begin{align}
\Psi(r_{*}) &\propto e^{-i\omega r_{*}}, \quad r_{*} \to -\infty, \\
\Psi(r_{*}) &\propto e^{+i\omega r_{*}}, \quad r_{*} \to +\infty.
\end{align}
These conditions uniquely define the quasinormal spectrum $\omega_{n\ell}$.

The Wentzel–Kramers–Brillouin (WKB) approach provides a semi-analytical approximation to the transmission and reflection coefficients of the wave equation. Matching the WKB solutions across the turning points near the peak of the potential leads to the condition \cite{Schutz:1985km,Iyer:1986np,Konoplya:2003ii}
\begin{equation}
\frac{i(\omega^2 - V_0)}{\sqrt{-2V_0''}} - \sum_{j=2}^{N} \Lambda_j = n + \tfrac{1}{2}, \quad n=0,1,2,\ldots,
\end{equation}
where $V_0$ and $V_0''$ are the value and the second derivative of the effective potential at its maximum, and $\Lambda_j$ are higher-order correction terms depending on higher derivatives of $V(r)$. In this work we use the sixth-order WKB expansion ($N=6$) \cite{Konoplya:2003ii}, which produces usually most accurate results in most cases, though the lower order formula when applied to $\ell >s$ perturbations, where $s$ is the spin of the field also provide reasonable accuracy \cite{Antonelli:2025yol,Pedrotti:2025idg,Konoplya:2020cbv,MahdavianYekta:2019pol,Matyjasek:2021xfg,Lutfuoglu:2025ljm,Konoplya:2023moy,Matyjasek:2017psv,Hamil:2025cms,Konoplya:2019hlu,Konoplya:2021ube,Paul:2023eep}, 
though the lower order formula when applied to $\ell >s$ perturbations, where $s$ is the spin of the field also provide reasonable accuracy \cite{Zhang:2025xqt,Miyachi:2025ptm,Konoplya:2005sy,Xiong:2021cth,Santos:2021wsw}.

The convergence of the WKB series can be significantly improved by applying Padé approximants \cite{Matyjasek:2017psv,Konoplya:2019hlu}. Instead of truncating the WKB expansion after the $N$-th term, one constructs a Padé approximant 
\begin{equation}
P^{\tilde m}_{\tilde n}(\varepsilon) = \frac{a_0 + a_1 \varepsilon + \ldots + a_{\tilde m} \varepsilon^{\tilde m}}
{1 + b_1 \varepsilon + \ldots + b_{\tilde n} \varepsilon^{\tilde n}},
\end{equation}
which reproduces the WKB series up to order $N = \tilde m + \tilde n$, with the bookkeeping parameter $\varepsilon$ set to unity at the end of the calculation.
 In practice, the choice $\tilde{m}=\tilde{n}=3$ yields the most stable and accurate results across a wide range of black-hole potentials \cite{Skvortsova:2024atk,Malik:2023bxc,Konoplya:2020jgt,Varghese:2011ku,Skvortsova:2023zmj,Dubinsky:2025fwv,Bolokhov:2024ixe,Bronnikov:2019sbx,Malik:2024nhy,Bolokhov:2023bwm,Skvortsova:2024wly,Konoplya:2019xmn,Lutfuoglu:2025hjy,Churilova:2021tgn,Dubinsky:2024nzo,Abdalla:2012si}. 

As an independent check, we compute the ringdown spectrum using numerical evolution in the time domain. We discretize the wave equation on a double-null grid according to the scheme of Gundlach, Price, and Pullin \cite{Gundlach:1993tp}:
\begin{equation}
\Psi(N) = \Psi(W) + \Psi(E) - \Psi(S) - \frac{\Delta^2}{8} \, V(S) \big( \Psi(W) + \Psi(E) \big),
\end{equation}
where $N, W, E, S$ denote the grid points at the future, west, east, and south corners of a null rectangle, and $\Delta$ is the grid step. Initial data are specified as a Gaussian pulse on an ingoing null surface. The resulting waveform shows the expected prompt response, followed by exponentially damped oscillations (the quasinormal ringing) and a late-time power-law tail.

To extract the quasinormal frequencies from the ringing stage, we employ the Prony method \cite{Prony:1795}, which fits the numerical signal to a superposition of damped exponentials,
\begin{equation}
\Psi(t) \approx \sum_{k=1}^{p} C_k \, e^{-i\omega_k t},
\end{equation}
where $C_k$ are complex amplitudes and $\omega_k$ are the complex frequencies. By choosing the fitting interval within the quasinormal ringing phase and adjusting the number of exponentials $p$, one can determine the dominant frequencies with high accuracy.

The combination of the WKB method with Padé resummation and the time-domain integration with Prony analysis provides a robust framework for computing quasinormal modes. The WKB approach yields accurate results for low-lying modes of moderate and high multipole numbers, while the time-domain method offers an independent confirmation and is particularly reliable when WKB accuracy decreases  \cite{Konoplya:2006gq}.

\begin{figure}
\resizebox{\linewidth}{!}{\includegraphics{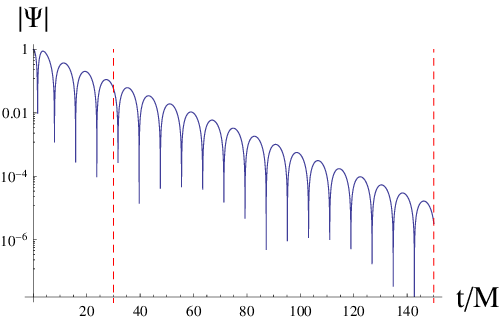}}
\caption{Semi-lograrithmic time-domain profile for $\gamma=1.18$, $\ell=2$, $M=1$. The Prony method gives $\omega = 0.3962 - 0.0742 i$, while the 6th order WKB produces $\omega = 0.396187-0.074213 i$. The red dashed lines enclose the time interval used for the Prony fitting.}\label{fig:TDprofile}
\end{figure}

\begin{table*}
\begin{tabular}{l c c c c}
\hline
\hline
$\gamma$ & WKB-6, $\widetilde{m}=4$ & WKB-8, $\widetilde{m}=4$ & rel. diff. $Re (\omega)$ & rel. diff. $Im (\omega)$  \\
\hline
$0$ & $0.373619-0.088929 i$ & $0.373669-0.088972 i$ & $0.0134\%$ & $0.0489\%$\\
$0.02$ & $0.373964-0.088779 i$ & $0.373993-0.088827 i$ & $0.00759\%$ & $0.0539\%$\\
$0.04$ & $0.374309-0.088638 i$ & $0.374318-0.088680 i$ & $0.00246\%$ & $0.0481\%$\\
$0.06$ & $0.374648-0.088498 i$ & $0.374645-0.088531 i$ & $0.00081\%$ & $0.0375\%$\\
$0.08$ & $0.374984-0.088356 i$ & $0.374975-0.088380 i$ & $0.00255\%$ & $0.0274\%$\\
$0.1$ & $0.375319-0.088210 i$ & $0.375306-0.088227 i$ & $0.00344\%$ & $0.0197\%$\\
$1.18$ & $0.396187-0.074213 i$ & $0.396195-0.074177 i$ & $0.00192\%$ & $0.0484\%$\\
\hline
\hline
\end{tabular}
\caption{Quasinormal modes ($n=0$) of the $\ell=2$,  perturbations calculated using the 6th and 8th order WKB formula with Padé  approximants for various values of $\gamma$; $M=1$.}
\end{table*}
\begin{table*}
\begin{tabular}{l c c c c}
\hline
\hline
$\gamma$ & WKB-6, $\widetilde{m}=4$ & WKB-8 $\widetilde{m}=4$ & rel. diff. $Re (\omega)$ & rel. diff. $Im (\omega)$  \\
\hline
$0$ & $0.599443-0.092703 i$ & $0.599443-0.092703 i$ & $0.00002\%$ & $0$\\
$0.02$ & $0.599922-0.092561 i$ & $0.599922-0.092561 i$ & $0.00001\%$ & $0$\\
$0.04$ & $0.600404-0.092417 i$ & $0.600404-0.092417 i$ & $0.00002\%$ & $0$\\
$0.06$ & $0.600889-0.092272 i$ & $0.600889-0.092272 i$ & $0.00001\%$ & $0$\\
$0.08$ & $0.601377-0.092124 i$ & $0.601377-0.092124 i$ & $0$ & $0.0001\%$\\
$0.1$ & $0.601869-0.091974 i$ & $0.601869-0.091974 i$ & $0$ & $0.0002\%$\\
$1.18$ & $0.634563-0.078131 i$ & $0.634564-0.078133 i$ & $0.00012\%$ & $0.0025\%$\\
\hline
\hline
\end{tabular}
\caption{Quasinormal modes ($n=0$) of the $\ell=3$ perturbations calculated using the 6th and 8th order WKB formula with Padé  approximants for various values of $\gamma$; $M=1$.}
\end{table*}
\begin{table*}
\begin{tabular}{l c c c c}
\hline
\hline
$\gamma$ & WKB-6, $\widetilde{m}=4$ & WKB-8, $\widetilde{m}=4$ & rel. diff. $Re (\omega)$ & rel. diff. $Im (\omega)$  \\
\hline
$0$ & $0.346466-0.273878 i$ & $0.346003-0.273555 i$ & $0.134\%$ & $0.118\%$\\
$0.02$ & $0.347072-0.273378 i$ & $0.346515-0.272975 i$ & $0.161\%$ & $0.147\%$\\
$0.04$ & $0.347679-0.272873 i$ & $0.347024-0.272300 i$ & $0.188\%$ & $0.210\%$\\
$0.06$ & $0.348285-0.272363 i$ & $0.347609-0.271405 i$ & $0.194\%$ & $0.352\%$\\
$0.08$ & $0.348891-0.271849 i$ & $0.348888-0.270209 i$ & $0.001\%$ & $0.604\%$\\
$0.1$ & $0.349496-0.271330 i$ & $0.350735-0.270779 i$ & $0.354\%$ & $0.203\%$\\
$1.18$ & $0.371000-0.227379 i$ & $0.370157-0.225937 i$ & $0.227\%$ & $0.634\%$\\
\hline
\hline
\end{tabular}
\caption{Quasinormal modes ($n=1$) of the $\ell=2$, perturbations calculated using the 6th and 8th order WKB formula with Padé  approximants for various values of $\gamma$; $M=1$.}
\end{table*}
\begin{table*}
\begin{tabular}{l c c c c}
\hline
\hline
$\gamma$ & WKB-6, $\widetilde{m}=4$ &  WKB-8, $\widetilde{m}=4$ & rel. diff. $Re (\omega)$ & rel. diff. $Im (\omega)$\\
\hline
$0$ & $0.582647-0.281310 i$ & $0.582645-0.281298 i$ & $0.00034\%$ & $0.00397\%$\\
$0.3$ & $0.591554-0.273843 i$ & $0.591548-0.273838 i$ & $0.00106\%$ & $0.00213\%$\\
$0.5$ & $0.597759-0.267795 i$ & $0.597767-0.267761 i$ & $0.00133\%$ & $0.0128\%$\\
$0.7$ & $0.604076-0.260578 i$ & $0.604113-0.260519 i$ & $0.00628\%$ & $0.0226\%$\\
$0.9$ & $0.610247-0.251818 i$ & $0.610307-0.251723 i$ & $0.00994\%$ & $0.0377\%$\\
$1.1$ & $0.615673-0.241109 i$ & $0.615708-0.240960 i$ & $0.00559\%$ & $0.0620\%$\\
$1.18$ & $0.617372-0.236237 i$ & $0.617399-0.236083 i$ & $0.00444\%$ & $0.0653\%$\\
\hline
\hline
\end{tabular}
\caption{Quasinormal modes ($n=1$) of the $\ell=3$ perturbations calculated using the 6th and 8th order WKB formula with Padé  approximants for various values of $\gamma$; $M=1$.}
\end{table*}
\begin{table*}
\begin{tabular}{c c c c c}
\hline
$\gamma $ & Prony fit & WKB-6, $\widetilde{m}=4$  & rel. diff. $Re (\omega)$ & rel. diff. $Im (\omega)$  \\
\hline
$0.1$ & $0.375316-0.088223 i$ & $0.375319-0.088210 i$ & $0.00076\%$ & $0.0148\%$\\
$0.35$ & $0.379646-0.086115 i$ & $0.379655-0.086105 i$ & $0.00232\%$ & $0.0111\%$\\
$0.6$ & $0.384318-0.083517 i$ & $0.384353-0.083503 i$ & $0.00912\%$ & $0.0173\%$\\
$0.85$ & $0.389335-0.080214 i$ & $0.389402-0.080289 i$ & $0.0173\%$ & $0.0935\%$\\
$1.1$ & $0.394567-0.075869 i$ & $0.394534-0.075908 i$ & $0.00843\%$ & $0.0521\%$\\
$1.18$ & $0.396224-0.074198 i$ & $0.396187-0.074213 i$ & $0.00925\%$ & $0.0202\%$\\
\hline
\end{tabular}
\caption{Comparison of the time-domain fit and the 6th order WKB formula with Padé  approximants for $\ell=2$ perturbations; $M=1$.}
\end{table*}
\begin{table*}
\begin{tabular}{c c c c c}
\hline
\hline
$\gamma$ & WKB-6, $\widetilde{m}=4$ & analytic & rel. diff. $Re (\omega)$ & rel. diff. $Im (\omega)$  \\
\hline
$0$ & $0.599443-0.092703 i$ & $0.602043-0.093011 i$ & $0.434\%$ & $0.332\%$\\
$0.02$ & $0.599922-0.092561 i$ & $0.602524-0.092870 i$ & $0.434\%$ & $0.334\%$\\
$0.04$ & $0.600404-0.092417 i$ & $0.603008-0.092727 i$ & $0.434\%$ & $0.335\%$\\
$0.06$ & $0.600889-0.092272 i$ & $0.603496-0.092583 i$ & $0.434\%$ & $0.337\%$\\
$0.08$ & $0.601377-0.092124 i$ & $0.603986-0.092436 i$ & $0.434\%$ & $0.339\%$\\
$0.1$ & $0.601869-0.091974 i$ & $0.604480-0.092287 i$ & $0.434\%$ & $0.341\%$\\
$1.18$ & $0.634563-0.078131 i$ & $0.637484-0.079229 i$ & $0.460\%$ & $1.41\%$\\
\hline
\hline
\end{tabular}
\caption{Quasinormal modes ($n=0$) of the $\ell=3$ perturbations calculated using the analytic formula and the 6th order WKB formula with Padé  approximanst for various values of $\gamma$; $M=1$.}
\end{table*}

\section{Quasinormal modes}

Although in this work the most accurate values of quasinormal frequencies are obtained numerically, it is instructive to begin with an approximate analytic treatment valid in the eikonal limit and its extensions. 

In a spherically symmetric background, perturbations reduce to a Schrödinger-like equation with an effective potential. In the large multipole limit, this potential can be expressed in the form
\begin{equation}\label{potential-multipole}
V(r_*) = \kappa^2 \left( H(r_*) + \Order{\kappa^{-1}} \right),
\end{equation}
where $\kappa \equiv \ell + \tfrac{1}{2}$. Following the expansion scheme developed in \cite{Konoplya:2023moy}, we treat the potential as a series in inverse powers of $\kappa$.

The function $H(r_*)$ has a single maximum, whose position can be represented as
\begin{equation}\label{rmax}
r_{\max} = r_0 + r_1 \kappa^{-1} + r_2 \kappa^{-2} + \ldots.
\end{equation}

Substituting (\ref{rmax}) into the first-order WKB approximation,
\begin{equation}
\omega = \sqrt{V_0 - i \K \sqrt{-2 V_2}},
\end{equation}
and then expanding consistently in powers of $\kappa^{-1}$, one arrives at the eikonal expression
\begin{equation}\label{eikonal-formulas}
\omega = \Omega \kappa - i \lambda \K + \Order{\kappa^{-1}}.
\end{equation}

Here $\Omega$ represents the angular frequency of the unstable null circular orbit (the photon sphere), while $\lambda$ is the corresponding Lyapunov exponent that characterizes the instability timescale of this orbit. Thus, in the eikonal regime the quasinormal spectrum is directly linked to the properties of the photon sphere, establishing the well-known correspondence between quasinormal modes and null geodesics \cite{Cardoso:2008bp}. The formula (\ref{eikonal-formulas}) provides a reliable approximation for the quasinormal frequencies in the regime $\kappa \gg \K$, i.e. for large multipole numbers.

While the geodesic correspondence offers a powerful analytic approximation in many cases, it is not universally valid. In \cite{Konoplya:2017wot,Khanna:2016yow} it was demonstrated that the relation breaks down when the standard centrifugal barrier term $f(r)\,\ell(\ell+1)/r^{2}$ in the effective potential is modified. Such deviations arise, for example, in a variety of higher-curvature theories \cite{Konoplya:2017wot,Konoplya:2020bxa,Konoplya:2019hml,Konoplya:2025afm,Bolokhov:2023dxq}. Another class of exceptions to the correspondence is connected with the inability of the WKB method to reproduce parts of the spectrum in the limit $\ell \rightarrow \infty$ \cite{Konoplya:2022gjp,Konoplya:2025mvj,Bolokhov:2023dxq}.

For the Hayward spacetime, however, the situation is more favorable. Provided that $\ell \geq 2$ and the quantum parameter $\gamma$ remains sufficiently small, the analytic eikonal formulas yield excellent agreement with numerical results, with relative errors typically below one percent.

Using the series expansion in powers of $\kappa\equiv\ell+1/2$ \cite{Konoplya:2023moy},  we find the position of a maximum of the effective potential,
\begin{widetext}
\begin{equation}
\begin{array}{rcl}
r_{\max } &=& \displaystyle\frac{M}{\kappa ^2}+3 M
\displaystyle+\gamma  \left(-\frac{4 M}{27 \kappa ^2}-\frac{2 M}{9}\right)
\displaystyle+\gamma ^2 \left(-\frac{17 M}{729 \kappa ^2}-\frac{M}{27}\right)
\displaystyle+\gamma ^3 \left(-\frac{130 M}{19683 \kappa ^2}-\frac{70 M}{6561}\right)\\
&&\displaystyle+\gamma ^4 \left(-\frac{1225 M}{531441 \kappa ^2}-\frac{665 M}{177147}\right)+\mathcal{O}\left(\gamma ^5,\frac{1}{\kappa ^4}\right),
\end{array}
\end{equation}
then, the 6th-order WKB formula for the quasinormal modes yields
\begin{equation}\label{eikonal}
\begin{array}{rcl}
\omega  &=& \displaystyle-\frac{i K \left(940 K^2-6599\right)}{46656 \sqrt{3} M \kappa ^2}-\frac{60 K^2+547}{1296 \sqrt{3} M \kappa }+\frac{\kappa }{3 \sqrt{3} M}-\frac{i K}{3 \sqrt{3} M}\\
&&\displaystyle+\gamma  \left(\frac{i K \left(820 K^2-833\right)}{104976 \sqrt{3} M \kappa ^2}+\frac{420 K^2-311}{34992 \sqrt{3} M \kappa }+\frac{\kappa }{81 \sqrt{3} M}+\frac{2 i K}{81 \sqrt{3} M}\right)\\
&&\displaystyle+\gamma ^2 \left(\frac{i K \left(2300 K^2-277\right)}{708588 \sqrt{3} M \kappa ^2}-\frac{3756 K^2+2735}{1889568 \sqrt{3} M \kappa }+\frac{\kappa }{486 \sqrt{3} M}+\frac{2 i K}{243 \sqrt{3} M}\right)\\
&&\displaystyle+\gamma ^3 \left(\frac{i K \left(331220 K^2+231047\right)}{306110016 \sqrt{3} M \kappa ^2}-\frac{124764 K^2+26911}{51018336 \sqrt{3} M \kappa }+\frac{61 \kappa }{118098 \sqrt{3} M}+\frac{181 i K}{59049 \sqrt{3} M}\right)\\
&&\displaystyle+\gamma ^4 \left(\frac{i K \left(2386436 K^2+8074523\right)}{16529940864 \sqrt{3} M \kappa ^2}-\frac{9311172 K^2+1438045}{5509980288 \sqrt{3} M \kappa }+\frac{2015 \kappa }{12754584 \sqrt{3} M}+\frac{3973 i K}{3188646 \sqrt{3} M}\right)+\mathcal{O}\left(\gamma ^5,\frac{1}{\kappa ^3}\right)
\end{array}
\end{equation}
\end{widetext}
where $\kappa\equiv\ell+1/2$ and $K\equiv n+1/2$.

The results summarized in Tables I–V allow us to draw two main conclusions concerning both the accuracy of the WKB method with Padé approximants and the influence of the quantum parameter $\gamma$ on the quasinormal spectrum. Table VI shows that the analytic approximation is reliable for $\ell=3$ and higher, with deviations remaining within one percent over most of the parameter range and exceeding this level only for near-extremal values of $\gamma$.

The comparison between the 6th- and 8th-order WKB expansions demonstrates that the Padé-improved WKB method is highly reliable in the present context. For the fundamental mode with $\ell=2$ (Table I), the relative differences between the two approximations are extremely small: the real part of the frequency differs by less than $0.02\%$, while the imaginary part shows deviations below $0.05\%$ across the whole range of $\gamma$. For $\ell=3$ (Table II), the agreement is even better, with differences of the order of $10^{-5}$–$10^{-4}\%$, which is essentially negligible compared to typical numerical uncertainties.  

For the first overtone ($n=1$), the situation is naturally more delicate, as higher overtones are less accurately captured by the WKB approximation. Nevertheless, for $\ell=2$ (Table III) the relative discrepancy between WKB-6 and WKB-8 remains moderate: $\lesssim 0.2\%$ for the real part and $\lesssim 0.6\%$ for the imaginary part. For $\ell=3$ and $n=1$ (Table IV), the relative errors stay comfortably below $0.07\%$ in the imaginary part and below $0.01\%$ in the real part. These values indicate that the WKB method, especially when Padé resummation is applied, retains high accuracy even for the first overtone.  

The reliability of the WKB approximation is further confirmed by direct comparison with time-domain simulations (Table V and fig. \ref{fig:TDprofile}). Here the Prony method was applied to extract the dominant frequencies from the ringdown waveform, and the results match the WKB predictions to within a fraction of a percent. Specifically, for $\ell=2$ and $\gamma$ ranging from $0.1$ to $1.18$, the relative differences in the real part of the frequency remain at the level of $0.001\%$, while the imaginary part differs by less than $0.1\%$. This remarkable agreement provides strong evidence that the WKB–Padé method offers a quantitatively precise description of the spectrum, fully consistent with independent time-domain analyses.  

The second important observation concerns the dependence of the spectrum on the quantum correction parameter $\gamma$. Across all multipoles and both fundamental and overtone modes, increasing $\gamma$ systematically shifts the spectrum toward larger oscillation frequencies and weaker damping. For instance, in the fundamental $\ell=2$ case (Table I), $Re(\omega)$ grows from $0.3736$ at $\gamma=0$ to $0.3962$ near the extremal value $\gamma=1.18$, corresponding to an increase of about $6\%$. Over the same range, the damping rate $|Im(\omega)|$ decreases from $0.0890$ to $0.0742$, i.e. by more than $15\%$. A similar trend is seen for $\ell=3$ (Table II), where the oscillation frequency rises by roughly $6\%$ and the damping rate decreases by about $16\%$ as $\gamma$ approaches extremality.  

For the first overtone, the effect of $\gamma$ is even more pronounced. In the $\ell=2, n=1$ case (Table III), $Re(\omega)$ increases from $0.346$ to about $0.371$ (a $\sim7\%$ growth), while the damping decreases from $0.274$ to $0.227$ (a $\sim17\%$ reduction). Similarly, for $\ell=3, n=1$ (Table IV), the oscillation frequency grows by approximately $6\%$ and the damping rate falls by almost $16\%$. This consistent pattern indicates that higher $\gamma$ values make the quasinormal ringing both faster and longer lived.  

These results can be physically understood in terms of the effective potential barrier. As $\gamma$ increases, the height of the potential grows while its width narrows, shifting the photon-sphere orbit to higher frequencies and reducing the leakage of waves through the barrier. The quasinormal spectrum therefore reflects this modification: larger $\gamma$ leads to higher oscillation frequencies and smaller imaginary parts. This behavior is in full agreement with the expectation that quantum corrections suppress dissipation and stabilize the oscillatory modes.

A striking feature of our results is that the influence of the quantum parameter $\gamma$ is considerably more pronounced for the first overtone than for the fundamental mode. This trend echoes the behavior identified for test fields in the Hayward spacetime \cite{Konoplya:2022hll}, where deviations from the Schwarzschild spectrum were observed to grow with the overtone number, leading to the phenomenon termed the “outburst of overtones” \cite{Konoplya:2022pbc}. Since overtones are more sensitive to the near-horizon geometry, their strong response can be interpreted as a direct imprint of quantum corrections that predominantly modify this region. Speculatively, this suggests that if future gravitational-wave observations were able to detect not only the dominant mode but also subdominant overtones with sufficient accuracy, they could serve as a particularly sensitive probe of near-horizon quantum structure. In this sense, the enhanced deviation of overtones may provide one of the most promising observational signatures of quantum-corrected black holes.

\section{Conclusions}

While quasinormal modes of test fields have been extensively investigated in the literature \cite{Toshmatov:2019gxg,Konoplya:2022hll,Lin:2013ofa,DuttaRoy:2022ytr}, the physically most important case of gravitational perturbations has not yet been analyzed in the context of quantum corrections arising from the Asymptotic Safety framework.  In this work we have carried out the first detailed study of gravitational quasinormal modes in the Hayward spacetime, which can be interpreted both as a regular black-hole model and as an effective quantum-corrected geometry arising in asymptotically safe gravity. Using the WKB method with Padé resummation and time-domain integration with Prony analysis, we computed the spectra of axial perturbations for the fundamental and first overtone, finding excellent agreement between the two approaches.  

In addition to the numerical results, in a similar fashion with \cite{Malik:2024wvs,Malik:2024bmp,Malik:2024sxv,Malik:2024qsz}, we derived an analytic expansion for gravitational quasinormal frequencies beyond the strict eikonal limit, making use of the $1/\kappa$ expansion of the effective potential. This approximation allowed us to verify the correspondence between eikonal quasinormal modes and the properties of unstable null geodesics. As expected, the relation holds in the large-$\ell$ regime, while deviations arise for smaller multipoles, where higher-order corrections become significant.  

Our numerical data show that the quantum parameter $\gamma$ systematically increases the oscillation frequency and decreases the damping rate, thereby producing faster and longer-lived ringdowns. The effect is particularly visible for the first overtone, in qualitative agreement with the phenomenon of the “outburst of overtones” reported earlier for test fields. At the same time, a definitive analysis of higher overtones requires more precise methods, such as Leaver’s continued-fraction approach, and is left for future work.  

Altogether, the Hayward spacetime provides a useful testbed for exploring the interplay between near-horizon quantum corrections and the quasinormal spectrum. The combination of analytic eikonal formulas, semi-analytical WKB estimates, and time-domain simulations offers a coherent picture of how quantum modifications alter the dynamics of black-hole perturbations. While present detectors are unlikely to access such fine details of the ringdown signal, next-generation gravitational-wave observatories may provide the necessary sensitivity to probe these effects observationally.

\begin{acknowledgments}
This work was supported by RUDN University research project FSSF-2023-0003.
\end{acknowledgments}

\bibliography{bibliography}

\end{document}